\documentstyle[titlepage,12pt]{article}
\input epsf

\def\dofig#1#2{\epsfysize=#1 \centerline{\epsfbox{#2}}}

\begin{document}

\thispagestyle{empty}
\begin{center}
{\Large\bf{Autonomous Traffic Signal Control Model with Neural
Network Analogy}}

\vspace{2em}
{\large{Toru Ohira}}

\vspace{2em}
Sony Computer Science Laboratory\\
3-14-13 Higashi-gotanda, Shinagawa, Tokyo 141, Japan\\
(SCSL-TR-97-001)\\
\end{center}

\renewcommand{\thesection}{\Roman{section}}

\begin{center}
\bf{Abstract}
\end{center}

We propose here an autonomous
traffic signal control model based on
analogy with neural networks. In this model,
the length of cycle time period of traffic 
lights at each signal is autonomously
adapted.  We find a self--organizing collective
behavior of such a model through simulation on
a one--dimensional lattice model road:
traffic congestion is greatly diffused when
traffic signals have such autonomous
adaptability with suitably tuned parameters. 
We also find that effectiveness of the system
emerges through interactions between
units and shows a threshold transition as a function of
proportion of adaptive signals in the model.   
\vspace{2em}

\centerline{\bf{1. INTRODUCTION}}
\vspace{1em}
The problem of traffic control is one of the major issues
of our society on motorways as well as 
computer networks. 
The problem has been of interest to both scientists and
engineers.
In the physics community, for example, the
main interest 
has been in investigating the nature of the
traffic itself.
Different approaches have been taken, and 
fluid dynamical models (Lighthill \& Whitham, 1955; Leutzbach, 1988), car--following models (Herman et al., 1959; Bando et al., 1995), 
cellular automaton models (Biham, 1992) and coupled lattice models 
(Yukawa \& Kikuchi, 1995) 
have been developed. Controlling of traffic has been
primarily an issue of the engineering community (See e.g., Huddart, 1996). 
Artificial neural network models have found many
engineering applications such as in optimizations, pattern 
recognitions and so on (See e.g., Arbib, 1995). 
We believe that the traffic control problem is one such area
that can benefit from ideas and knowledge from neural network
research, particularly with a diffusion of ideas
stemming from other fields mentioned above.
With this motivation, we propose here an autonomous
signal control model which is based on a cellular automaton
traffic modeling
approach with an adaptive signal control mechanism in analogy with
neural networks.
The collective behavior of the model is investigated by
computer simulations, and it is found that in a certain
parameter range the adaptive model can diffuse a
traffic jam. We also found that the effectiveness of
the system is due to interactions between these adaptive
signal units rather than as a sum of separate contributions
from individual units. This is inferred from the fact 
that the effectiveness of the model shows a phase
transition as a function of proportion of adaptive signals
in the system.
Even though our model shows an acceptable performance within a certain 
range of parameters, we are not aiming for a good 
engineering system to solve the traffic problem here.
The main focus of this paper, rather, is to present a model 
designed at a cross section of the above mentioned fields so 
that it can 
serve as a possible starting point of exploration for diffusion of a variety of 
knowledge developed within each research area on the matter of traffic. 

\vspace{1.5em}
\centerline{\bf{2. AUTONOMOUS SIGNAL CONTROL MODEL}}
\vspace{1em}

Let us now describe our model in more detail. For simplicity
we restrict ourselves here to model a one dimensional 
road with closed boundary condition (loop) and discrete
time and step motions of cars and signals.
Extension to other geometry, however, is
straightforward and will be discussed elsewhere. 
The model road has $N$ discrete sites.
At every $D$ distance there is a signal with
$U$ signals in total. 
The $i$th signal is characterized by 
two parameters, $T^i_b$ and $T^i_r$, which are the duration of blue
and red lights, respectively. The ``cycle length'' of
the signal is defined by
\begin{equation}
T^i = T^i_b + T^i_r
\end{equation}
For simplicity we take all the signals to have an equal ``split'' between 
blue and red: $T^i_b = T^i_r$ and $T_i= 2*T^i_b$. 
We introduce $M$ number of cars all of which go along the model one--way
road in the same direction. U-turns and passing are not allowed.
A car can move forward a unit step per unit time 
if the position in front of it is 
not occupied by another car. If the next position is the signal position,
it can move when both of the following two conditions are satisfied:
the signal is in the blue state and the position just beyond the signal is
vacant (i.e., cars cannot be at the signal position). 

The main algorithm of the 
model is described by the following signal dynamics:
\begin{equation}
T^i_b(t) =  \phi(V^i(t)), \quad
V^i(t+1)  =  {X^i_b(t)}.
\end{equation}
$V^i(t)$ is a ``potential'' 
of $i$th signal at $t$th period
 and duration of the blue signal
is given by a bounded non-linear mono--tone increasing function $\phi$,
which is taken as a sigmoidal shape:
\begin{equation}
\phi(u) = {T_0} + \gamma \tanh(\beta u), \quad (\gamma < {T_0}),
\end{equation}
where $T_0$, $\beta$ and $\gamma$ are parameters.
$X^i_b(t)$  
is the 
number of cars reaching the signal when the signal light is blue, 
i.e., the number of cars going through the
signal position at $t$th period of blue light. (When a car cannot
move beyond the signal because of a jam, we include it
in the count.)
Because of the above condition of equal split between blue and red lights,
this dynamics, in effect,  changes the signal cycle length $T^i$. 
The dynamics of both cars and signals are synchronous: all
cars and signals are updated at each step according to the above transition 
rule.
The dynamics of signal cycle in this model
has a natural correspondence with the neural network models. 
Each signal is identified by an integrate--and--fire neuron (Farley \& Clark, 1961; Beurle, 1962; Milton et al., 1993) and the 
traffic going through a signal is identified by the neural 
pulses it receives. The activity level of
the $i$th neuron is identified by $T^i_b$ which is a bounded nonlinear
function of potential $V^i$. After each period (or firing), 
the potential $V^i$ is reset to zero
as in neuron models.  This analogy motivates us to look into
the possibility of emergent effective collective behaviors of these adaptive
units, which have been observed with neural network models.
In the following we show that this is indeed the case, and
that the adaptive signals with the above
dynamics collectively lead to the 
overall diffusion of the traffic congestion. 

\vspace{1.5em}
\centerline{\bf{3. SIMULATION EXPERIMENTS}}
\vspace{1em}

The traffic congestion is
quantified by measuring average velocity, $\langle v(k) \rangle$.
In our model, we define $\langle v(k) \rangle$ as an ensemble
average of the proportion of cars which moved at time
step $k$. More precisely, $\langle v(k) \rangle = \langle L(k)/M \rangle$,
where $L(k)$ is the number of cars moved at time step $k$.
When we start with random initial positions of cars and
all signals with $T^i_b = T^i_r = T_0$ but with different states and
phases, the $T^i_b$ starts to change and shows non-homogeneity. Typically, the
average signal period $\langle T_b \rangle$ over all signals settles
to a stationary value after a transient period. 
The average velocity of cars $\langle v(k) \rangle$ also leads into an oscillating 
stationary state.
Figure 1 shows one such typical example. In the figure we also compared
the model with the case of all signals have a fixed period at $\langle T_b \rangle$.
We observe that the adaptive signal model can retain 
higher collective velocity and less congestion after a transient
period in this example.

\begin{figure}[h]
\dofig{3.3in}{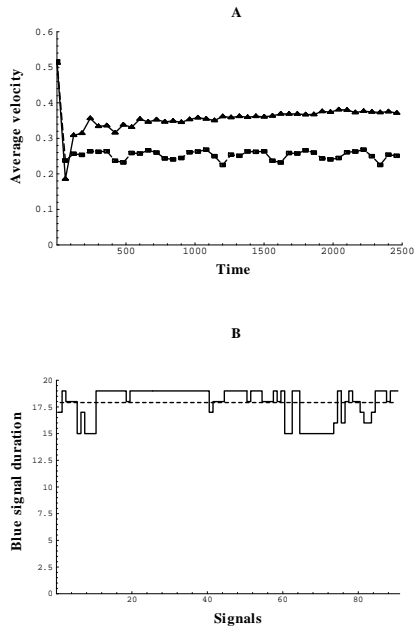}
\caption{
One comparison  
between the adaptive model
and fixed signal period model. 
(A)
The dynamics of average velocity $\langle v(k) \rangle$
is compared between the adaptive model (triangles) and fixed
period model (squares) with $\langle T^i_b \rangle$. 
The parameters are set as $N=450$, $M = 150$, $D=4$, $U=90$,
$T_0=10$, $\gamma=9$, $\beta = 0.16$, 
and $\langle T^i_b \rangle$=18.
Each trial started with random initial positions of cars and phase of
signals
and was averaged over 100 trials.
(B) Distribution of blue signal duration $T^i_b$ at
2500 steps appeared in one of the trials. The dashed
line represent $\langle T^i_b \rangle \approx 18$.
}
\end{figure}

This observation also holds  
with different car densities and parameter settings when $\langle T_b \rangle$ is
sufficiently larger than inter-signal distance $D$.
Some results are shown in Figure 2. Up to $\approx$50$\%$ improvement
has been observed. Qualitatively similar results are obtained
for various system sizes of $N = $ 100, 200, and 450.

\begin{figure}[h]
\dofig{3.5in}{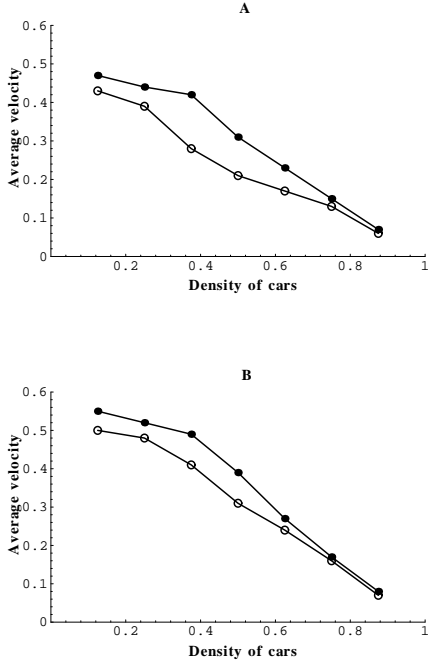}
\caption{
Examples of comparison between the
adaptive (black circle) and fixed (open circle) signal period models
as the density of cars $M/(N-U)$ in the system is varied,
with $N=100$, $D=5$, and $U=20$.
The parameters are set as 
(A) $T_0 = 7$, $\gamma=6$,   
and $\langle T^i_b \rangle$ = 12.
(B) $T_0 = 10$, $\gamma=9$,   
and $\langle T^i_b \rangle$ = 18.
$\beta$ for the adaptive models 
is tuned for each $M$.
Qualitatively similar results are also
obtained for $N=200,450$.
}
\end{figure}

We infer from these results that the model here with suitably tuned
parameters can 
show emergent collective organization of signal period distribution
by gradually adapting to particular traffic 
conditions on the model road which are otherwise homogeneous.

To gain more insight into the collective behavior of the model, we investigate
how performance of the model changes when we
stop the adaptation after some initial period and the signal
periods are thereafter fixed. 
Figure 3 shows an example of the result from such simulations: there is
a sharp drop in average velocity after the adaptation is stopped. This
tells us that adaptation dynamics at each signal is crucial in keeping
average velocity high, rather than a particular configuration of 
period distributions over all the signals.

\begin{figure}[h]
\dofig{2.5in}{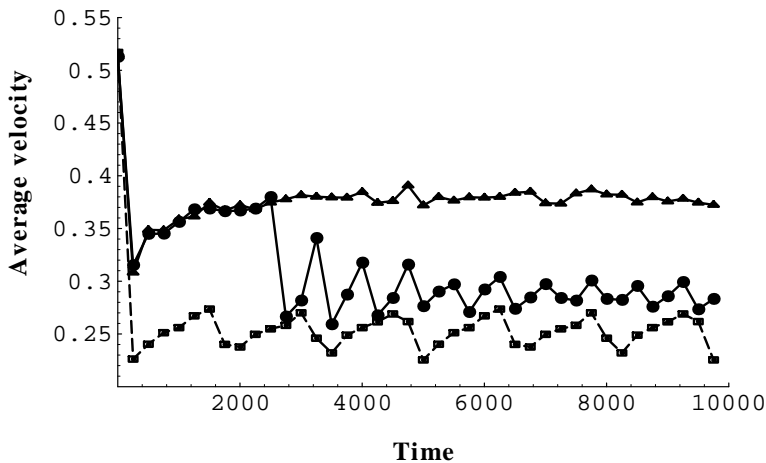}
\caption{
Example of comparison of dynamics of the 
average velocity. 
The parameters are set as $N=450$, $M = 150$,
$T_0=10$, $\gamma=9$, $\beta = 0.16$, 
and $\langle T^i_b \rangle$=18.
They are adaptive model (triangles), adaptive model 
changed to fixed cycle model at step 2500 (circles), and
fixed cycle model (squares). 
}
\end{figure}
  
We also looked at the case where only some portion of signals have
the adaptive capability and others operate at a fixed period; 
a representative example is shown in Figure 4. We see that the
performance of the system measured in the average velocity is
quite non--linear and threshold--like as a function of
the proportion of adaptive signals.

This suggests that the collective behavior of the model is
not simply an aggregation of the effect of individual signals.
Rather, the interaction among signals, which is indirectly
mediated by cars passing through, is playing a role in the
collective behavior of the model system. 
\clearpage

\begin{figure}[h]
\dofig{2.5in}{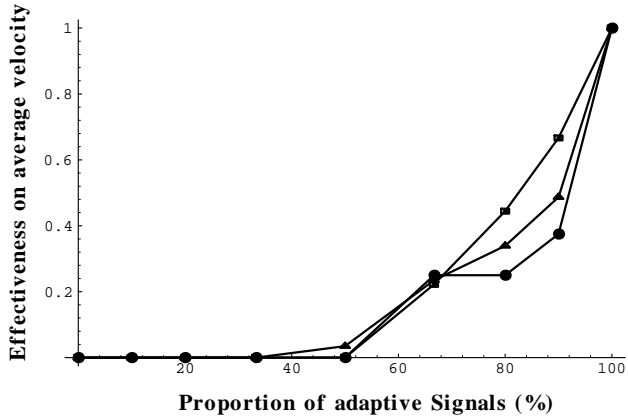}
\caption{
Change of effectiveness of average velocity
as a function of propotion of the number of adaptive
signal units. The effectiveness is measured
in average velocity at 2500 steps and rescaled
between (0,1) from no adaptive element
to all adaptive elements.
The parameters are set as $N=450$, $D=5$, $U=90$,
$T_0=10$, $\gamma=9$, $\beta = 0.16$, 
and $\langle T^i_b \rangle$=18.
The number of cars are varied $M=100$ (squares),
$150$ (triangles), and $200$ (circles). 
Qualitatively similar results are also
obtained for $N=100, 200$.
}
\end{figure}

\vspace{1.5em}
\centerline{\bf{4. CONCLUSION}}
\vspace{1em}

The adaptive model presented here belongs to a class of models
often termed emergent computation models (Huberman, 1992; Forrest, 1991). 
These models
aim to derive effective computation out of interacting autonomous 
local actions from each unit in the system, rather than by
top--down style algorithms or by a central control of units. 
Even though the guiding principles of designing such models vary,
experience, concepts, and  insight gained from studies of systems showing
emergent behaviors as in neural network modelings can be
quite useful particularly for the theoretical understanding
of the model, which is left for the future.
It is hoped that our model with the neural network analogy presented here
can serve to call for attention to looking at traffic control problems
as another research area of neural network community.

\begin{center}
{\bf REFEREENCES}
\end{center}
\vspace{1em}

\noindent
Arbib, L. (1995). The Handbook of Brain Theory and Neural Networks.
Cambridge, MA: MIT Press.
\vspace{1em}

\noindent
Bando, M., Hasebe, K., Nakayama, A., Shibata, A., \& Sugiyama, Y. (1995).
Dynamics model of traffic congestion and numerical simulation.
 {\it Physical Review E},
{\bf 51}, 1035--1042.
\vspace{1em}

\noindent
Beurle, R. L. (1962). Functional organization in random networks.
In H. v. Foerster \& G. W. Zopf, (Eds.), {\it Principles of Self-Organization}
(pp. 291--314). New York: Pergamon Press.
\vspace{1em}

\noindent
Biham, O., Middleton, A. A., \& Levine, D. (1992). 
Self--organizetion and a dynamical transition in traffic--flow models.
{\it Physical Review A},
{\bf 46}, R6124--6127.
\vspace{1em}

\noindent
Farley, B. G. \&  Clark, W. A. (1961). Activity in networks of neuron--like
elements. In C. Cherry, (Ed.), {\it Information Theory}
(pp. 242--251).
Washington: Butterworths.
\vspace{1em}

\noindent
Forrest, S. (1991). Emergent Computation.
Cambridge, MA: MIT Press.
\vspace{1em}

\noindent
Herman, R., Montroll, E. W., Potts, R. B., \& Rothery, R. W. (1959).
Traffic dynamics: Analysis of stability in car following.
 {\it Operations Research},
{\bf 7}, 86--106.
\vspace{1em}

\noindent
Huberman, B. A. (1992). Computation: The Micro and The Macro View.
Singapore: World Scientific.
\vspace{1em}

\noindent
Huddart, K. W.(chair) (1996). {\it Eigth International Conference on Road Traffic Monitoring and Control}, London: IEE Conference Publication 422.
\vspace{1em}

\noindent
Leutzbach, W. (1988). Introduction to the Theory of Traffic Flow.
Berlin: Springer--Verlag.
\vspace{1em}

\noindent
Lighthill, M. J. \& Whitham, G. B. (1955).
On kinetic waves II A theory of trraffic flow on long crowded roads. {\it Proceedings of Royal Society A},
{\bf 229}, 317--345.
\vspace{1em}

\noindent
Milton, J. G., Chu, P. H., \& Cowan, J. D. (1993). Spiral waves in integrate--and--fire neural networks. In S. J. Hanson, J. D. Cowan, \& C. L. Giles (Eds.),
{\it Neural Information Processing Systems}(pp. 1001--1006). 
San Mateo: Morgan Kaufmann.
\vspace{1em}

\noindent
Yukawa, S. \&  Kikuchi, M. (1995). 
Coupled--map modeling of one--dimensional traffic flow.
{\it Journal of the Phyical Society of Japan}.
{\bf 64}, 35--38.
\vspace{1em}

\end{document}